# Membrane Interactions in Alzheimer's Treatment Strategies with Multitarget Molecules

Pablo Zambrano[1*]

[1]Department of Bioscience, School of Natural Sciences, Technical University of Munich, Lichtenbergstrasse 4, 85748 Garching, Germany.

*Corresponding author: pablo.zambrano@tum.de

The neuronal membrane is one of the cell structures most affected in neurodegenerative diseases. The harmful and toxic effects that specific molecules related to the etiology of Alzheimer's disease have on the plasma membrane of neurons have been widely demonstrated. With the increasing amount of experimental data regarding the malfunctioning of proteins (folding and accumulation), many new therapies have been proposed that could eventually mitigate the effect of such molecules (for example, tau protein and beta-amyloid peptide). However, most related research has focused on the direct action of targets with specific neuronal functions (for example, NMDA receptor antagonists, AChE blockers and amyloid peptide aggregates). Recently, a comprehensive review of emergent multitarget strategies in *Bioorganic Chemistry* was performed by Pathak et al. (2024) [1], who explored the complexities of the multifaceted nature of Alzheimer's disease and the myriad of therapeutic targets. Their exhaustive discourse provides a pivotal understanding of how these novel multitarget agents could coordinate several therapeutic effects beyond the classical targets. Although these strategies are indispensable, the impact of molecules on the phospholipid membrane of neurons could be equally significant, an aspect that has usually been overlooked. These little-explored aspects not only limit the complete understanding of the mechanism of action of new molecules but also the understanding of the role that the lipid membrane plays in the pathogenesis of neurodegenerative diseases. Therefore, exploring the protective role of these new agents on the neuronal plasma membrane could lead to the development of new therapeutic approaches and a more integrated view of the development of these pathologies. In this letter, I briefly detail some examples and offer new ideas for addressing this approach.

Neurodegenerative diseases represent a growing threat to global public health, with Alzheimer's disease (AD) being one of the most prevalent and devastating. AD is characterised by a progressive loss of memory and other cognitive functions, leading to a decline in quality of life and, eventually, death. The central pathology of AD includes the accumulation of beta-amyloid peptide (Aβ) plaques in the brain, resulting in significant neurotoxicity and subsequent neuronal death. Current treatments for AD are primarily palliative and have limited efficacy in modifying disease progression [2]. Moreover, side effects associated with these treatments often restrict their use in patients. Therefore, there is an urgent need to develop more effective and safer therapies for AD and other neurodegenerative diseases. In this context, multitarget-directed ligands (MTDLs) have emerged as an innovative option for addressing the complexity of AD [3]. These molecules are designed to interact with multiple relevant biological targets capable of modulating different aspects of the disease, such as neuroinflammation, oxidation, and Aβ aggregation. Research into MTDLs, such as donepezil-huprine [4] (Fig. 1A), rhein-huprine [5] (Fig. 1B) or capsaicin-tacrine hybrids [6] (Fig. 1C), has yielded encouraging results in both in vitro and in vivo models, revealing new perspectives for AD treatment. These molecules have proven effective in modulating Aβ toxicity and have provided exciting insights into the interaction between MTDLs and cellular membrane models. The meticulous design of these molecules, grounded in a deep understanding of the molecular biology of AD, has been driven by technological advances in molecular modelling and biochemical analysis. Specifically, rhein-huprine and donepezil-huprine hybrids have proven promising for modulating Aβ(1-42) toxicity and interactions with cellular membranes, thus providing potential safeguards against Aβ-induced neurotoxicity. On the other hand, capsaicin-tacrine hybrids have structural similarities to previous hybrids (Figure 1), and their ability to interact with membrane models has been preliminarily studied by a parallel artificial membrane permeability assay (PAMPA) [6].

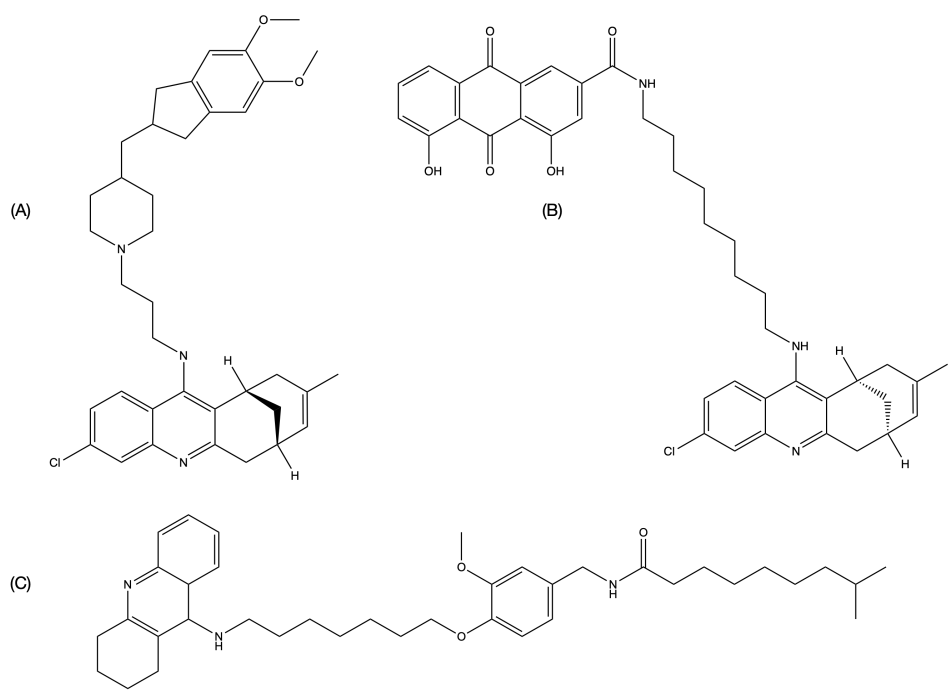

**Figure 1.** Representative structures of the (A) donepezil-huprine hybrid [AVCRI104P4], (B) rhein-huprine hybrid, and (C) capsaicin-tacrine hybrid.

The implications of these findings are vast. First, they further validated the multitarget approach for modulating AD pathogenesis. Moreover, evidence of the effective interaction of these molecules with cellular membrane models and their ability to mitigate Aβ(1-42) toxicity provides a solid foundation for further exploration of MTDLs in in vivo models and eventually in clinical trials. Notably, many therapeutic strategies currently do not consider the lipid membrane as a target or, at least, as an adjuvant to new therapies for neurodegenerative diseases. This omission might obscure crucial aspects in understanding and treating these diseases, suggesting that further exploration of the interactions of therapeutic molecules with lipid membranes might be beneficial.

While some of these molecules, such as rhein-huprine and donepezil-huprine hybrids, can insert into the lipid membrane and protect it against the disruptive effect of the Aβ(1-42) peptide, there might be other mechanisms by which these molecules exert their effects at a more macroscopic level (such as through in vivo experiments). Could these molecules also interact with structures that enhance the excretion or clearance of the Aβ(1-42) peptide from nerve cells? If this is not the case, how could this capability be improved? The behaviour of huprine hybrids on lipid bilayers also suggests a potential mechanism of action in modulating the function and structure of

ABC transporters present in neuronal cell membranes. This interaction could influence the activity of these transporters, potentially enhancing amyloid clearance. The ABCA7 transporter, for which extensive information regarding its function and structure is available [7], has high homology with other ABC transporters involved in lipid transport; therefore, it is plausible that by interacting with the lipid membrane, huprine and tacrine hybrids might also influence the function of these related transporters, which could have implications for lipid transport dynamics and the formation and removal of amyloid plaques.

On the other hand, the ability of these hybrids to stabilise the neuronal membrane indirectly enhances the efficiency of ABC transporters in removing amyloid from the brain, as an intact neuronal membrane is essential for correct activity. Additionally, from a pathogenic point of view, since ABCA7 can transfer lipids from cell membranes to lipoproteins [8], tacrine and huprine hybrids might influence this process, impacting neuronal lipid homeostasis and potentially AD pathogenesis. Ultimately, by interacting with the membrane and ABC transporters beyond amyloid clearance, these hybrids might modulate other cellular processes related to the disease, such as neuroinflammation and the response to oxidative stress.

Another target of interest for future multitarget strategies is the lysosomal system. Given that lysosomal dysfunction, as recently shown in a study involving chloroquine [9], can lead to the intracellular accumulation of Aβ(1-42), it is reasonable to speculate on the therapeutic potential of molecules that can act on the lysosomal system to combat the pathogenesis of AD. MTDLs that can insert themselves into a lipidic membrane and protect it from the attack of Aβ(1-42) might offer a novel strategy. These molecules can act not only as protectors of the membrane but also as modulators of lysosomal function, enhancing the degradative activity of enzymes such as cathepsin D (CTSD) and therefore reducing the intracellular accumulation of Aβ(1-42) [10]. Moreover, by improving the integrity of the membrane, huprine hybrids might also strengthen the blood–brain barrier, which plays a crucial role in the pathogenesis of AD. Together, the combination of membrane protection, the improvement in amyloid clearance through ABC transporters, and the modulation of lysosomal function could offer multiple therapeutic approaches to combat AD, addressing the extracellular and intracellular accumulation of Aβ(1-42). However, further in-depth research is needed to fully understand the possible mechanism of action of these molecules and their therapeutic potential in in vivo and in vitro models.

The exploration of MTDLs is an exciting strategy in the search for more effective therapies for AD and other neurodegenerative conditions. However, as in the search for new therapeutic agents, optimising artificial membrane models is equally important. Designing membrane models similar to natural neuronal membranes would open a vast field of study for therapeutic molecules or physiological phenomena that are not fully understood. In this sense, emerging efforts in synthetic biology are auspicious. Although the creation of a synthetic cell from scratch is still limited, studies of membrane reconstitution from the bottom up could offer numerous advantages for in vitro studies in neurosciences, especially for diseases where the plasma membrane plays a crucial role.

In summary, detailed studies on the interactions of these molecules with synthetic models of lipidic membranes and their ability to attenuate the toxicity of the beta-amyloid peptide open a wide range of possibilities for the design of new drugs. Ultimately, the development of MTDLs for neuroprotection represents a significant step toward achieving more effective and safer therapies for neurodegenerative diseases and underscores the importance of a holistic approach in the fight against these devastating conditions.